\documentclass[twocolumn,showpacs,floatfix,superscriptaddress,prl]{revtex4}
\usepackage{amsmath} \usepackage{amssymb} \usepackage{graphicx}
\usepackage{dcolumn} \usepackage{bm}

\begin{document}
\title{Universal Distribution of Kondo Temperatures in Dirty Metals}
\author{P. S. Cornaglia}
\author{D. R. Grempel} 
\affiliation{CEA-Saclay, DSM/DRECAM/SPCSI,
B\^at. 462, F-91191 Gif-sur-Yvette, France}
\author{C. A. Balseiro}
\affiliation{Centro At\'{o}mico Bariloche, 8400 San
Carlos de Bariloche, R\'{\i}o Negro, Argentina}

\begin{abstract}
Kondo screening of diluted magnetic impurities in a disordered
host is studied analytically and numerically in one, two and three
dimensions. It is shown that in the $T_{\textsc k} \to 0$ limit
the distribution of Kondo temperatures has a universal form,
$P(T_{\textsc k}) \sim T_{\textsc k}^{-\alpha}$ that holds in the 
insulating phase and persists in the metallic phase close to the 
metal insulator transition. Moreover, the exponent $\alpha$
depends only on the dimensionality. The most important consequence
of this result is that the $T$ dependence of thermodynamic
properties is smooth across the metal-insulator transition in
three dimensional systems.
\end{abstract}


\maketitle

In a class of compounds with coexisting magnetic moments and
conduction electrons the magnetic susceptibility and the specific
heat coefficient exhibit anomalous power law or logarithmic low
temperature divergences. However, magnetic order is not observed
down to the lowest attainable temperatures \cite{Miranda2005}.
This behavior is observed in a growing family of heavy-fermion
compounds \cite{Steward2001} an also in the doped semiconductor
Si:P \cite{Paalanen1991}.

It was early recognized that this type of non-Fermi liquid (NFL)
anomaly is associated with the interplay of the Kondo effect and
strong disorder. In heavy fermions disorder is present mostly in
the $f$-electron sub-system. In a clean system screening of
$f$-moments starts at the single-impurity Kondo scale $T_{\textsc
k} \sim \exp[- 1/(J \rho_0)]$ where $J$ is the Kondo coupling and
$\rho_0$ is conduction electron density of states (DOS) at the
Fermi level. In the dirty case $J$ is a random variable with a
distribution $P(J)$. Therefore, the local values of $T_{\textsc
k}$ are also random and have a distribution $P(T_{\textsc k})$
solely determined by $P(J)$. The simple Kondo-disorder model
\cite{Bernal1995} and its subsequent improvements
\cite{Miranda1996}, based on the above physical picture, lead
quite naturally to NFL phenomenology.

One might think that when the disorder is mostly present in the
{\it conduction band} $P(T_{\textsc k})$ is similarly determined
by $\rho({\bf R},\varepsilon_{\textsc f})$, the distribution of
the local DOS (LDOS) of the host at the Fermi level at the
position of the Kondo impurity
\cite{Dobrosavljevic1992}.

In this Letter we show that this is
not the case except at weak disorder. In general, $P(T_{\textsc
k})$ depends on subtle features of the LDOS and its statistical
properties, not simply on the distribution of its values at the
Fermi energy \cite{Dobrosavljevic1992b,Aguiar2003,Kaul2005}. 
The most important consequence
of the analysis presented below is that in the limit $T_{\textsc
k} \to 0$, the form of $P(T_{\textsc k})$ is {\it universal} in 
the insulating phase. We find that
$P(T_{\textsc k}) \sim A\;T_{\textsc k}^{-\alpha}$ where $A$ is a
non-universal amplitude and the exponent $\alpha$ depends only on
the dimensionality. Furthermore, the same behavior is obtained 
for the metallic phase close to the metal insulator transition,
with an identical exponent.

We discuss the problem of a single magnetic impurity Kondo-coupled
to a disordered bath of non-interacting electrons. We represent
the host by an Anderson model with random site energies uniformly
distributed in an interval $[-W/2,W/2]$. Without loss of
generality we shall set the chemical potential to zero and place
the impurity at {\bf R}=0. We shall denote the dimensionality of
the system by $D$ ( $D$ = $1$, $2$ and $3$). The host is an insulator for
$W > W_c$ where $W_c$ is finite in three dimensions (3D) and $W_c =
0$ in one (1D) and two dimensions (2D).

Following previous authors \cite{Miranda1996,Dobrosavljevic1992}
we use slave-boson mean field theory to write down an implicit
equation for $T_{\textsc k}$
\begin{eqnarray}\label{eq:largeN0}
\frac{2}{J}&=&\int_{-\infty}^\infty \frac{d\varepsilon} {
\varepsilon}\; \tanh(\varepsilon/2 T_{\textsc
k})\;\rho(\varepsilon,{\bf 0})\;.
\end{eqnarray}
Here, $\rho(\varepsilon,{\bf 0})=\sum_{\ell} \delta(\varepsilon
-\varepsilon_\ell)\;|\psi_\ell({\bf 0})|^2$ is the LDOS of the 
host at ${\bf R}=0$, where $\psi_\ell({\bf R})$ is
 the $\ell$-th eigenfunction of
the host and $\varepsilon_\ell$ is its associated eigenvalue for
a fixed realization of the disorder. This formula is identical to 
what is obtained defining $T_{\textsc k}$ as the temperature at which second 
order perturbation in $J$ breaks down \cite{Nagaoka}.

In the first part of the Letter we discuss extensively the
properties of $P(T_{\textsc k})$ in the Anderson insulator for
which a number of results can be obtained using physical arguments
and the solving analytically a simplified model that captures
the essential physics. Then, we present the results of numerical
simulations for all the insulating phases and for the metallic
phase in 3D.

In an Anderson insulator the wave functions are localized around
random positions {\bf R}$_\ell$. At long distances from their peak
positions they decay as $|\psi_\ell({\bf R} - {\bf R}_\ell)|^2
\sim \exp(- 2 |{\bf R} - {\bf R}_\ell|/\xi_{\ell})$ where
$\xi_{\ell}\equiv \xi(\varepsilon_\ell)$ is the energy-dependent
localization length. Since $\xi(\varepsilon)$ is smooth on the scale
of $T_{\textsc k}$   we shall neglect its energy dependence and
denote its value at $\varepsilon=0$ by $\xi$.  The eigenvalues
$\varepsilon_\ell$ are random variables and their distribution is
Poissonian~\cite{Mott-book}.

For a fixed configuration of the site energies the LDOS consists
of a dense set of $\delta$-functions with varying amplitudes.
There are ${\cal O}(\xi^D)$ peaks with weights ${\cal
O}(\xi^{-D})$ that come from states localized within a volume of
size $~ \xi^D$ around the impurity and infinitely many small peaks
with exponentially smaller weights that come from states localized
outside this volume. The mean separation between the main peaks is
$\Delta E_\xi \sim B(W)/\xi^D$ where $B(W)$ is the
bandwidth ({\it i.e.} the support of the LDOS).

Consider first the contribution of the largest peaks to the LDOS
and neglect the rest. The conditions of validity of this
approximation
 will be discussed below. It is convenient to split the interval
of integration in Eq.~(\ref{eq:largeN0}) in subintervals of width
$T_{\textsc k} << B(W)$. Within each of these the function
$\tanh(\varepsilon/2T_{\textsc k})/\varepsilon$ is essentially
constant and can be safely taken out of the sub-integral.
Then, the LDOS in
Eq.~(\ref{eq:largeN0})
can be well approximated by the coarsegrained function
$\rho_{s}(\varepsilon) \sim  N_p(\varepsilon,T_{\textsc
k})/(\xi^D T_{\textsc k})$ where $  N_p(\varepsilon,T_{\textsc
k})$ is the number of large peaks with energies in an interval of
width $T_{\textsc k}$ around $\varepsilon$. It is clear that $
\overline{N_p(\varepsilon,T_{\textsc k})} = \xi^D T_{\textsc
k}/B(W)$ and $\overline{\rho}_s = B^{-1}(W)$. The LDOS may be
split in two parts, $\rho_{s}(\varepsilon)= \overline{\rho}_s +
\delta \rho_{s}(\varepsilon)$ where  $\delta
\rho_{s}(\varepsilon)$  is a random function with zero mean.

Suppose that $\overline{N_p} \gg 1$. Then, $\delta
\rho_{s}(\varepsilon)$ is a Gaussian random function with
$\overline{\delta \rho_{s}(\varepsilon)\;\;\delta
\rho_{s}(\varepsilon^\prime)} =
\overline{\rho}_s\;\xi^{-D}\;\delta(\varepsilon -
\varepsilon^\prime)$. Equation~(\ref{eq:largeN0}) and
elementary properties of Gaussian random functions lead readily to
\begin{eqnarray}\label{eq:gaussian}
P(T_{\textsc k}) \propto \exp\left[- \frac{\ln^2 \left(T_{\textsc
k}/T^0_{\textsc k}\right)}{ 2 (\xi_{\textsc k}/\xi)^D
 }\right].
\end{eqnarray}
This is a log-normal distribution centered at $T^0_{\textsc k}$,
the Kondo temperature
 of a clean system with a uniform
LDOS $\overline{\rho}_s$. The variance of the distribution is
$\sigma^2=(\xi_{\textsc k}/\xi)^{D}$ where $\xi^D_{\textsc k}=
1/(T^0_{\textsc k} \rho_0)$ is the volume of the Kondo screening
cloud~\cite{Nozieres1974}. Since $\sigma^{-2} \sim
\overline{N_p}$, Eq.~(\ref{eq:gaussian}) is valid for $\xi
\gtrsim \xi_{\textsc k}$. In this regime most of the impurities
have Kondo temperatures in a narrow interval around $T^0_{\textsc
k}$ and yet we are in an insulating state.

This is not the end of the story. Even for $\xi
> \xi_{\textsc k}$ the condition $\overline{N_p} \gg 1$ is
 violated for $T_{\textsc k}\lesssim B(W)/\;\xi^D$.
 For these temperatures, the number of large peaks
with energies in an interval $T_{\textsc k}$ around the Fermi
energy is small. They no longer dominate the LDOS as assumed and
screening of the impurity must be done by electron states
localized far away from the former. But these are precisely those
that we neglected from the start. We then expect that for $\xi >
\xi_{\textsc k}$ $P(T_{\textsc k})$ will show, in addition of the
peak around $T^0_{\textsc k}$, a low temperature tail appearing
below $T_{\textsc k}\;\sim B(W)/\;\xi^D$ carrying a small fraction
of the weight of the former. For $\xi < \xi_{\textsc k}$ there is
no temperature range in which Eq.~(\ref{eq:gaussian}) holds and we
do not expect to find a sharp peak near $T^0_{\textsc k}$.  The
tail should still
appear below $\sim T^0_{\textsc k}$.

In order to gain insights on the properties of the low-temperature
tail in $P(T_{\textsc k})$ we now discuss an exactly solvable
``toy'' model for the insulating phase that reveals interesting
features that we shall retrieve in the numerical solution of the
Anderson model presented below.

As
stated above the localized functions $\psi_\ell({\bf R})$ decay
exponentially away from their peak positions. However, they
fluctuate randomly around the envelope $\sim \exp(- R/\xi)$. In
our toy model these fluctuations are ignored and the true
eigenfunctions are replaced by normalized exponentials centered
around their random locations. Furthermore, these positions ${\bf
R}_{\ell}$ are assumed to be uncorrelated with the energies. This
holds deep in the insulating phase. Equation~(\ref{eq:largeN0})
becomes
\begin{equation}\label{largeN3}
\frac{2}{J}= \sum_{\ell} \frac{e^{-2 R_\ell /\xi}} {{\cal
N}_D} \frac{\tanh(\varepsilon_\ell/2T)}{\varepsilon_\ell}
\equiv F_T(\{\varepsilon_{\ell}\})\;,
\end{equation}
where  the normalization ${\cal N}_D=2^{-D}\Omega_D \xi^D (D-1)!$,
$\Omega_D$ is the solid angle in D dimensions and the last term in
the above equation defines the random function $F_T$.

A key feature of Eq.~(\ref{largeN3}) is that $F_T$ is a
monotonically decreasing function of $T$. A moment's reflection
will convince the reader that this implies
\begin{eqnarray}
P(F_T < 2/J) = P(T_{\textsc k} < T) + P(T_{\textsc k} = 0)\;,
\end{eqnarray}
where the first term on the right hand side is the integrated
probability of $T_{\textsc k}$  
from zero to $T$, and the second is the probability of the
impurity remaining unscreened at $T=0$. Therefore, the
distribution of {\it non-zero} Kondo temperatures is given by
\begin{eqnarray}\label{eq:Ptkdef}
P(T_{\textsc k})=\int_0^{2/J}\left. \frac{\partial
W_T(x)}{\partial T}\right|_{T=T_{\textsc k }} dx,
\end{eqnarray}
where  $W_T(x) =
\overline{\delta\left(F_T\left(\{\varepsilon_{\ell}\}\right) -
x\right)}$ is the probability density of $F_T$. In our toy model
$W_T(x)$ can be calculated analytically leading to a closed form
expression for $P(T_{\textsc k})$. 

Here, we are interested in the asymptotic properties
 in three relevant regimes. For  $T_{\textsc
k}/W \gg 1/\mathcal{N_D}\sim \xi^{-D}$ we find that $P(T_{\textsc k})$ 
sharply decays for large
$T_{\textsc k}$ as $\sim \exp(-C \xi^D T_{\textsc
k}\ln^2T_{\textsc k} )$ where $C$ is a constant of order one. The
same expansion is valid for $\xi
> \xi_{\textsc k}$, $T_{\textsc k}\sim T_{\textsc k}^0$ and in this case we
retrieve precisely Eq.~(\ref{eq:gaussian}).

More importantly, we can now have access to the low-temperature
region that was beyond the scope of the qualitative analysis
presented above. In the limit $T \to 0$ we find
\begin{equation} \label{eq:PtktoylowT}
P(T_{\textsc k})\sim \Omega_D\left(\frac{\xi}{2}\right)^D|\log(T_{\textsc
k}/W)|^\alpha\;\;\;{\textrm {for}}\;\; T_{\textsc k}\to 0\;,
\end{equation}
with $\alpha = D - 1$.
The asymptotic form of $P(T_{\textsc k})$ for
this model is thus  a power of the logarithm with an exponent
depending only on dimensionality. The prefactor, proportional to
$\xi^D$, decreases with increasing disorder. A more detailed
analysis shows that this divergent contribution appears at very
low temperatures for $\xi\gg \xi_{\textsc k}$. With increasing
disorder the tail shifts to higher temperatures.

The model discussed above is based on a crude representation of
the localized states. Furthermore, it can obviously not describe
the metallic phase in 3D or the metal-insulator transition (MIT). 
Therefore, we turn to our numerical results for the Anderson model (AM).
It will be seen that the two models share the same {\it
qualitative} properties. Namely, a divergent $P(T_{\textsc k})$  
at low $T_{\textsc k}$ with
 the nature of the divergence depending only the spatial 
dimension and not on disorder. We shall also see that 
$P(T_{\textsc k})$
acquires the {\it same} form in the localized and
metallic phases in 3D.

Our numerical calculations were performed using a representation
of the host by the Anderson model on a regular lattice of size
$L^D$ for $D$ = 1, 2, and 3:
\begin{equation} \label{eq:Anderson}
H_A = \sum_{i,\sigma}\;\varepsilon_i c^{\dagger}_{i \sigma}
c^{}_{i \sigma}  -t \sum_{\langle i,j \rangle,\sigma}\;
c^{\dagger}_{i \sigma} c^{}_{j \sigma}\;, \end{equation} where 
$\varepsilon_i$ are random variables and the hopping integral
between nearest neighbors $t=1$ is 
the unit of energy. We 
show results for $J =1/4 B$, where $B$ is the bandwidth of the
clean system, and for $\varepsilon_i$ uniformly distributed in the
interval $[-W/2,W/2]$. We have checked the robustness of our
results, in particular the value of the exponent of the power law,
analyzing other values of $J$ and considering a Gaussian
distribution for $\varepsilon_i$.

For each lattice size and each realization of the random potential
the Anderson Hamiltonian was diagonalized and
Eq.~(\ref{eq:largeN0}) was solved for $T_{\textsc k}$ using the
numerical eigenfunctions and eigenvalues. Histograms were
constructed repeating the procedure for many realizations of the
potential and positions of the impurity. The size of the lattice
and the number of realizations was increased until convergence of
the histograms. The values of the parameters 
appear in the captions of the figures.

\begin{figure}[htbp]
\includegraphics[width=8.5cm,clip=true]{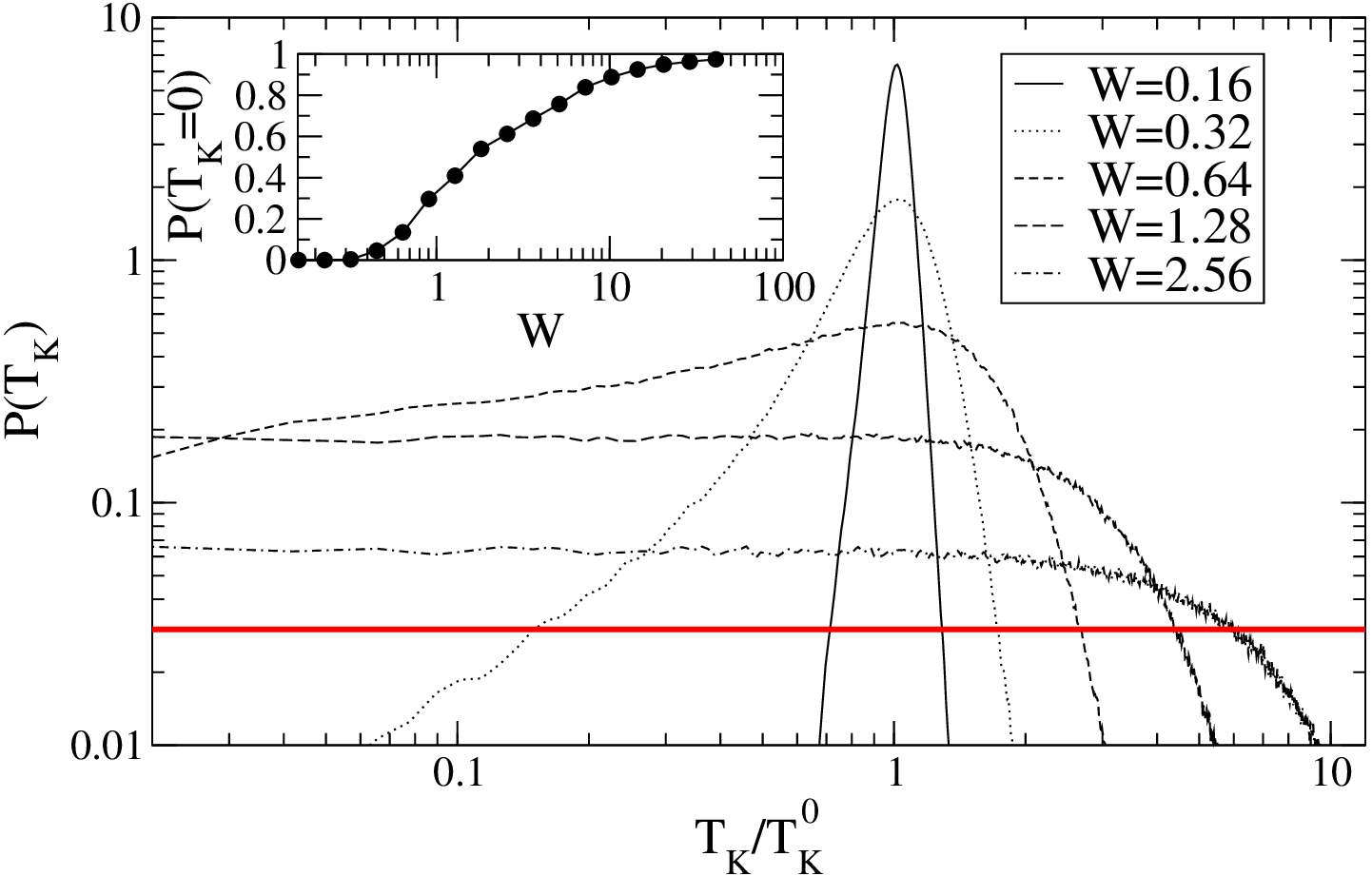}
\caption{$P(T_{\textsc k})$ in 1D for several values of the
amplitude of the disorder. Results for a chain of $N_s=1000$
sites. The number of realizations of the potential is $1000$.
Inset: Fraction of unscreened impurities at $T=0$ as a function of
disorder.} \label{fig:fig1}
\end{figure}

Figure \ref{fig:fig1} displays the distribution of Kondo
temperatures in $D=1$. It is seen that at small disorder
$P(T_{\textsc k})$ is strongly peaked at $T_{\textsc k}^0$. For $W
\ll 1$ the shape of the peak fits accurately
Eq.~(\ref{eq:gaussian}).
With increasing disorder $P(T_{\textsc k})$ first broadens and
departs from the log-normal distribution, and a tail appears at
low temperatures. Upon increasing further the disorder, the peak
at $T_{\textsc k} = T_{\textsc k}^0$ disappears and $P(T_{\textsc
k})$ becomes flat as in the toy model.

\begin{figure}[htbp]
\includegraphics[width=8.5cm,clip=true]{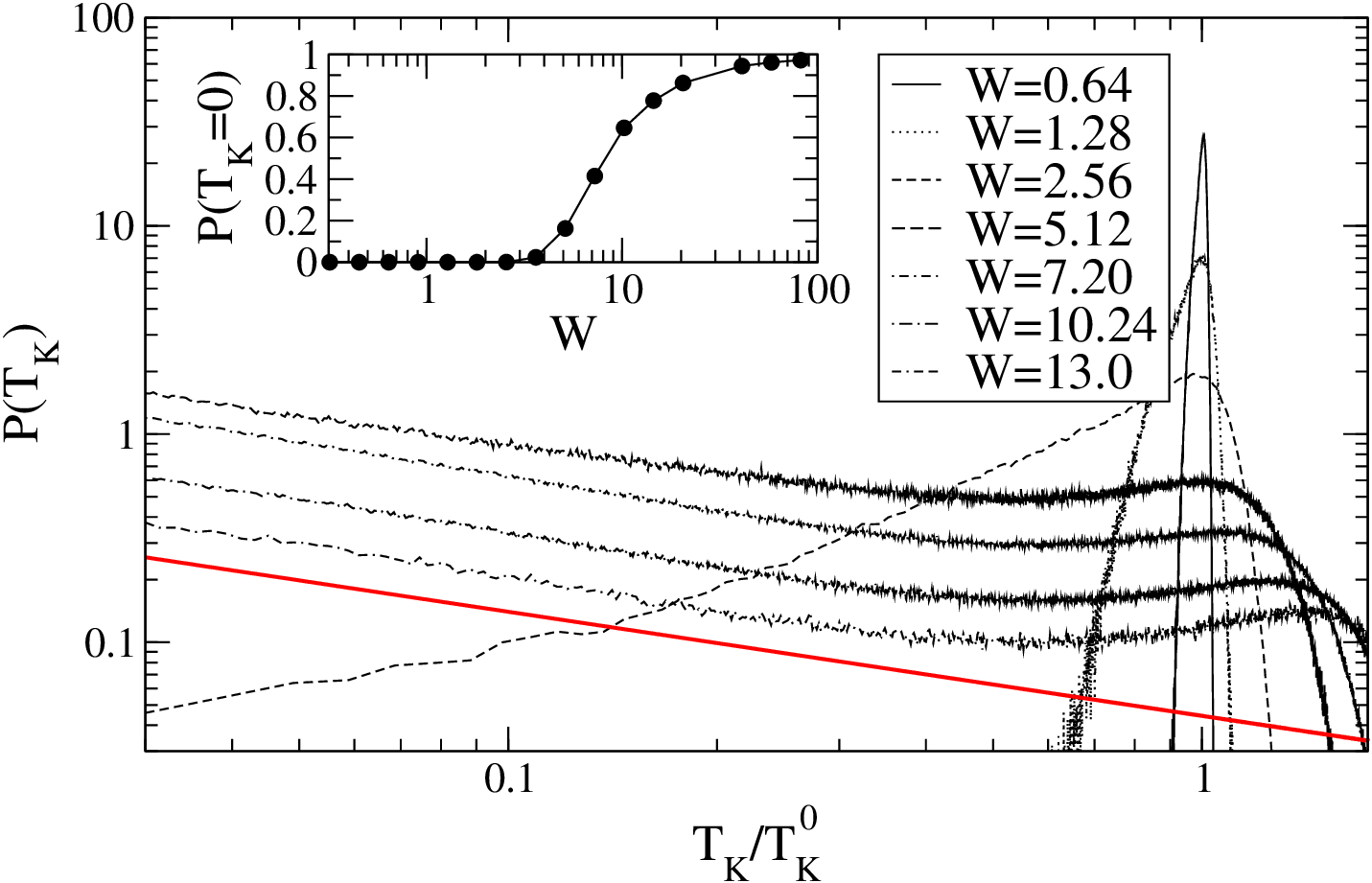}
\caption{$P(T_{\textsc k})$ in 2D for several values of the
amplitude of the disorder. Results for square lattices of
$N_s=2025$ sites and 1000 realizations of the amplitude of the
potential. The slope of the thick straight line is -0.5. Inset:
Fraction of unscreened impurities at $T=0$ as a function of
disorder.} \label{fig:fig2}
\end{figure}

The fraction of unscreened impurities at $T=0$ as a function of
disorder is shown in the inset to Fig. \ref{fig:fig1}.
$P(T_{\textsc k}=0)$ increases continuously with disorder from
vanishingly small values for $W\sim 0$ to ${\cal O}(1)$ for large
$W$. 
We independently
estimated $\xi$ as a function of disorder and checked that the
crossover between the two regimes takes place when $\xi\sim
\xi_{\textsc k}$,
as expected from our qualitative arguments.

The results for the two dimensional case are displayed in
Fig.~\ref{fig:fig2}.
The same general trends are also observed in
this case but in two dimensions $P(T_{\textsc k})$ is no longer
flat but divergent at low $T_{\textsc k}$ as expected from our 
analytical results.
Note that the straight lines corresponding to different values 
of $W$ are
parallel meaning that $P(T_{\textsc k})$ is a power law with a
{\it disorder  independent} exponent. The thick straight line has slope $-
1/2$ suggesting that $\alpha = 1/2$ in 2D.

For temperatures below the range plotted in the figure finite-size
effects come into play. 
These are expected 
for $T_{\textsc k} \lesssim \Delta$, the typical level spacing 
at the Fermi level $\propto
L^{-D}$~\cite{Thimm1999}.
 We have thus performed a careful
study of these effects and checked that the power law behavior
extends to lower and lower values of $T_{\textsc k}$ as the size
of the system increases.

\begin{figure}[htbp]
\includegraphics[width=8.5cm,clip=true]{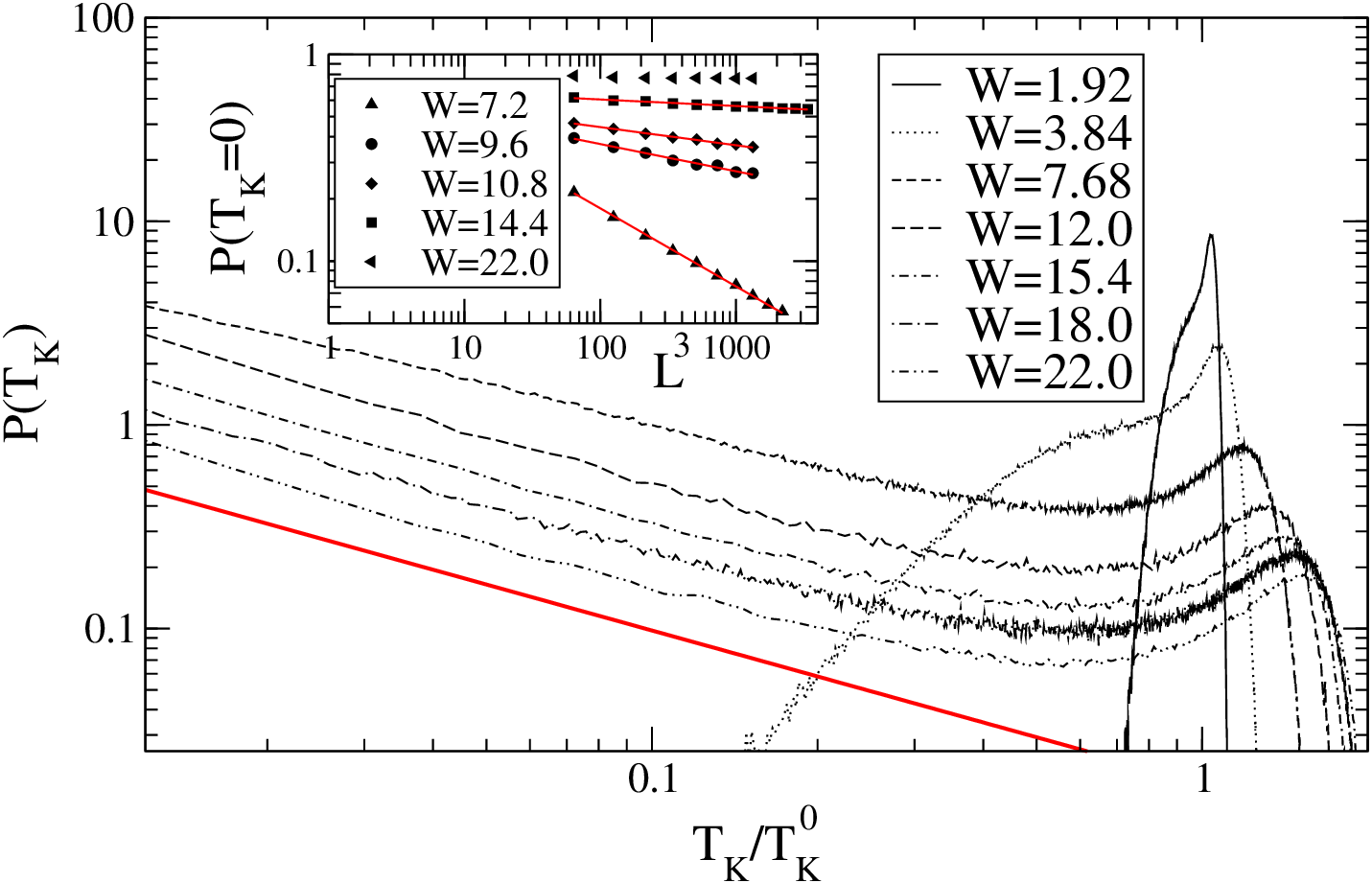}
\caption{$P(T_{\textsc k})$ in 3D for several values of the
amplitude of the disorder. Results for cubic lattices of
$N_s=2197$ sites and $1000$ realizations of the amplitude of the
potential. The MIT occurs at $W_c \sim 16.5$. The slope of the
thick straight line is -0.75. Inset: size and disorder dependence
of $P(T_{\textsc k}=0)$ in the metallic phase.
}
\label{fig:fig3}
\end{figure} 

The results for the 3D case, displayed in Fig.~{\ref{fig:fig3}},
are the most interesting ones because in this case there
is a MIT at $W_c \sim 16.5$ \cite{Mackinnon1981}. As in the 2D
case, $P(T_{\textsc k})$ is a power law at low temperature but the
slope of the thick straight line is now $-3/4$ suggesting that
this is the value of the exponent in 3D. A striking new feature
 in 3D is that the
form of $P(T_{\textsc k})$ and the exponent are {\it the same} on
both sides of the MIT. This can be seen by comparing the two
lowermost curves in Fig.~\ref{fig:fig3}, that correspond to
$W>W_c$, to the two above them that correspond to $W<W_c$.
The total contribution of the impurities to the susceptibility 
 or specific heat can be calculated using interpolation formulas 
(see e.g. Ref. \cite{Miranda2001}) and weighting the individual contributions with
P($T_K$).  An important consequence is that the magnetic susceptibility and the
specific heat coefficient in this model should exhibit a power law
behavior with a disorder independent exponent, $\chi\;\sim\;\gamma\;\sim\;T^{-\alpha}$, and be
continuous across the MIT. Interestingly, this is
precisely what was observed in Si:P which has a MIT as a 
function of the concentration of P and shows signatures of local magnetic 
moments in both the metallic and insulating phases \cite{Paalanen1991,Hirsch1992}.
Although the agreement obtained is very appealing, this system contains physical ingredients 
which are out of the scope of our model. 

We found that in our finite systems a fraction of the spins
remains unscreened down to zero temperature. This fraction,  very
small for small disorder, increases as $W \to W_c$. Since on
general grounds, in the thermodynamic limit, all spins must be
screened at $T=0$  in the metallic phase, the observed effect is
due to finite-size. The inset to Fig. \ref{fig:fig3} displays the
results of a finite-size scaling analysis of $P(T_{\textsc k}=0)$
for several values of $W$. It can be seen
that $P(T_{\textsc k}=0)$ scales to zero in the metallic phase 
with increasing $L$ as it
should. Moreover, $P(T_{\textsc k}=0) \propto L^{-\beta(W)}$ with
an exponent $\beta(W)$ that depends on the disorder and that
decreases as $W$ increases. In contrast, deep in the insulating
phase, $P(T_{\textsc k}=0)$ remains finite in the thermodynamic
limit \cite{Dobrosavljevic1992b}.

In this Letter we have studied the single-impurity problem and our
results are strictly applicable in the diluted limit. We showed that
 the structure of the LDOS induces a power
 law $P(T_{\textsc k})$ in the insulating phase. This low $T_{\textsc k}$ 
divergence is associated to impurities screened by states localized 
at distances larger the localization length $\xi$. The statistical 
properties of these {\it far} states are such that a change 
in the disorder strength shifts the onset temperature of the diverging behavior but not
 the nature of the divergence. Surprisingly, the exponent of the power 
law is disorder independent and in 3D the same behavior persists across the MIT with an identical exponent.

In concentrated systems interaction effects can not be ignored. An
extension of the Dynamical Mean Field Theory (DMFT) method for 
random systems was used with
success in that context~\cite{Aguiar2003,Dobrosavljevic1997}. 
A power law $P(T_{\textsc k})$ was found
for the disordered Anderson model on a Bethe lattice using DMFT in
Ref.~\cite{Miranda2001}.  An important difference between this
work and ours is that the exponent of the power law found in
Ref.~\cite{Miranda2001} depends on the strength of disorder and is
therefore non-universal. It would be very interesting to
investigate how our results for the diluted case evolve with
increasing density and what is the role of the
lattice geometry.

The authors thank C. Castellani, C. Di Castro, V. Dobrosavljevi{\'c},
M. Grilli, V. Kravtsov, and E. Miranda for
illuminating discussions. Part of this work was performed at the
University of Rome La Sapienza. P.S.C. and D.R.G. thank the
Istituto dei Sistemi Complessi (CNR) and the MIUR COFIN 2003 grant
No 2003020230-006 for financial support. This work was partially
supported by Fundaci\'on Antorchas, grant No. 14248/21.

After the completion of this work we became aware of a related study by
S. Kettemann and E. R. Mucciolo \cite{Ketteman2005}.


\begin{thebibliography}{20}
\expandafter\ifx\csname natexlab\endcsname\relax\def\natexlab#1{#1}\fi
\expandafter\ifx\csname bibnamefont\endcsname\relax
  \def\bibnamefont#1{#1}\fi
\expandafter\ifx\csname bibfnamefont\endcsname\relax
  \def\bibfnamefont#1{#1}\fi
\expandafter\ifx\csname citenamefont\endcsname\relax
  \def\citenamefont#1{#1}\fi
\expandafter\ifx\csname url\endcsname\relax
  \def\url#1{\texttt{#1}}\fi
\expandafter\ifx\csname urlprefix\endcsname\relax\def\urlprefix{URL }\fi
\providecommand{\bibinfo}[2]{#2}
\providecommand{\eprint}[2][]{\url{#2}}

\bibitem[{\citenamefont{Miranda and Dobrosavljevi{\'c}}()}]{Miranda2005}
\bibinfo{author}{\bibfnamefont{E.}~\bibnamefont{Miranda}} \bibnamefont{and}
  \bibinfo{author}{\bibfnamefont{V.}~\bibnamefont{Dobrosavljevi{\'c}}},
  \bibinfo{journal}{Rep. Prog. Phys.} \textbf{\bibinfo{volume}{68}},
  \bibinfo{pages}{2337} (\bibinfo{year}{2005}).

\bibitem[{\citenamefont{Steward}(2001)}]{Steward2001}
\bibinfo{author}{\bibfnamefont{G.}~\bibnamefont{Steward}},
  \bibinfo{journal}{Rev. Mod. Phys.} \textbf{\bibinfo{volume}{73}},
  \bibinfo{pages}{797} (\bibinfo{year}{2001}).

\bibitem[{\citenamefont{Paalanen and Bhatt}(1991)}]{Paalanen1991}
\bibinfo{author}{\bibfnamefont{M.~A.} \bibnamefont{Paalanen}} \bibnamefont{and}
  \bibinfo{author}{\bibfnamefont{R.~N.} \bibnamefont{Bhatt}},
  \bibinfo{journal}{Physica B} \textbf{\bibinfo{volume}{169}},
  \bibinfo{pages}{231} (\bibinfo{year}{1991});
\bibinfo{author}{\bibfnamefont{H.}~\bibnamefont{v.~Lohneysen}}, in
  \emph{\bibinfo{booktitle}{Adv. in Sol. State Phys.}}
  (\bibinfo{publisher}{Vieweg Braunschweig/Wiesbaden}, \bibinfo{year}{2000}),
  vol.~\bibinfo{volume}{40}, p. \bibinfo{pages}{143}.

\bibitem[{\citenamefont{Bernal et~al.}(1995)\citenamefont{Bernal, MacLaughlin,
  Lukefahr, and Andraka}}]{Bernal1995}
\bibinfo{author}{\bibfnamefont{O.~O.} \bibnamefont{Bernal}},
  \bibinfo{author}{\bibfnamefont{D.~E.} \bibnamefont{MacLaughlin}},
  \bibinfo{author}{\bibfnamefont{H.~G.} \bibnamefont{Lukefahr}},
  \bibnamefont{and} \bibinfo{author}{\bibfnamefont{B.}~\bibnamefont{Andraka}},
  \bibinfo{journal}{Phys. Rev. Lett.} \textbf{\bibinfo{volume}{75}},
  \bibinfo{pages}{2023} (\bibinfo{year}{1995}).

\bibitem[{\citenamefont{Miranda et~al.}(1996)\citenamefont{Miranda,
  Dobrosavljevi{\'c}, and Kotliar}}]{Miranda1996}
\bibinfo{author}{\bibfnamefont{E.}~\bibnamefont{Miranda}},
  \bibinfo{author}{\bibfnamefont{V.}~\bibnamefont{Dobrosavljevi{\'c}}},
  \bibnamefont{and} \bibinfo{author}{\bibfnamefont{G.}~\bibnamefont{Kotliar}},
  \bibinfo{journal}{J. Phys.:Condens. Matter} \textbf{\bibinfo{volume}{8}},
  \bibinfo{pages}{9871} (\bibinfo{year}{1996});
 \bibinfo{author}{\textit{ibid.}},
  \bibinfo{journal}{Phys. Rev. Lett.} \textbf{\bibinfo{volume}{78}},
  \bibinfo{pages}{290} (\bibinfo{year}{1997}).

\bibitem[{\citenamefont{Dobrosavljevi{\'c}
  et~al.}(1992)\citenamefont{Dobrosavljevi{\'c}, Kirkpatrick, and
  Kotliar}}]{Dobrosavljevic1992}
\bibinfo{author}{\bibfnamefont{V.}~\bibnamefont{Dobrosavljevi{\'c}}},
  \bibinfo{author}{\bibfnamefont{T.~R.} \bibnamefont{Kirkpatrick}},
  \bibnamefont{and} \bibinfo{author}{\bibfnamefont{G.}
  \bibnamefont{Kotliar}}, \bibinfo{journal}{Phys. Rev. Lett.}
  \textbf{\bibinfo{volume}{69}}, \bibinfo{pages}{1113} (\bibinfo{year}{1992}).

\bibitem[{\citenamefont{Dobrosavljevic and
  Kotliar}(1992)}]{Dobrosavljevic1992b}
\bibinfo{author}{\bibfnamefont{V.}~\bibnamefont{Dobrosavljevic}}
  \bibnamefont{and} \bibinfo{author}{\bibfnamefont{G.}~\bibnamefont{Kotliar}},
  \bibinfo{journal}{Phys. Rev. B} \textbf{\bibinfo{volume}{46}},
  \bibinfo{pages}{5366} (\bibinfo{year}{1992}).

\bibitem[{\citenamefont{Aguiar et~al.}(2003)\citenamefont{Aguiar, Miranda, and
  Dobrosavljevi{\'c}}}]{Aguiar2003}
\bibinfo{author}{\bibfnamefont{M.~C.~O.} \bibnamefont{Aguiar}},
  \bibinfo{author}{\bibfnamefont{E.}~\bibnamefont{Miranda}}, \bibnamefont{and}
  \bibinfo{author}{\bibfnamefont{V.}~\bibnamefont{Dobrosavljevi{\'c}}},
  \bibinfo{journal}{Phys. Rev. B} \textbf{\bibinfo{volume}{68}},
  \bibinfo{eid}{125104} (\bibinfo{year}{2003}).

\bibitem[{\citenamefont{Kaul et~al.}(2005)\citenamefont{Kaul, Ullmo,
  Chandrasekharan, and Baranger}}]{Kaul2005}
\bibinfo{author}{\bibfnamefont{R.~K.} \bibnamefont{Kaul}},
  \bibinfo{author}{\bibfnamefont{D.}~\bibnamefont{Ullmo}},
  \bibinfo{author}{\bibfnamefont{S.}~\bibnamefont{Chandrasekharan}},
  \bibnamefont{and} \bibinfo{author}{\bibfnamefont{H.~U.}
  \bibnamefont{Baranger}},
  \bibinfo{journal}{Europhys. Lett.}
  \textbf{\bibinfo{volume}{71}}, \bibinfo{pages}{973}
  (\bibinfo{year}{2005}).

\bibitem[{\citenamefont{Nagaoka}(1965)\citenamefont{Nagaoka}}]{Nagaoka}
\bibinfo{author}{\bibfnamefont{Y.} \bibnamefont{Nagaoka}},
  \bibinfo{journal}{Phys. Rev}
  \textbf{\bibinfo{volume}{138}}, \bibinfo{pages}{A1112}
  (\bibinfo{year}{1965}).

\bibitem[{\citenamefont{Mott}(1990)}]{Mott-book}
\bibinfo{author}{\bibfnamefont{N.~F.} \bibnamefont{Mott}},
  \emph{\bibinfo{title}{Metal-Insulator transition}}
  (\bibinfo{publisher}{Taylor and Francis, London}, \bibinfo{year}{1990}).

\bibitem[{\citenamefont{Nozi{\`e}res}(1974)}]{Nozieres1974}
\bibinfo{author}{\bibfnamefont{P.}~\bibnamefont{Nozi{\`e}res}},
  \bibinfo{journal}{J. Low Temp. Phys.} \textbf{\bibinfo{volume}{17}},
  \bibinfo{pages}{31} (\bibinfo{year}{1974}).

\bibitem[{\citenamefont{Thimm et~al.}(1999)\citenamefont{Thimm, Kroha, and von
  Delft}}]{Thimm1999}
\bibinfo{author}{\bibfnamefont{W.~B.} \bibnamefont{Thimm}},
  \bibinfo{author}{\bibfnamefont{J.}~\bibnamefont{Kroha}}, \bibnamefont{and}
  \bibinfo{author}{\bibfnamefont{J.}~\bibnamefont{von Delft}},
  \bibinfo{journal}{Phys. Rev. Lett.} \textbf{\bibinfo{volume}{82}},
  \bibinfo{pages}{2143} (\bibinfo{year}{1999});
  \bibinfo{author}{\bibfnamefont{I.} \bibnamefont{Affleck}} \bibnamefont{and}
  \bibinfo{author}{\bibfnamefont{P.} \bibnamefont{Simon}},
  \bibinfo{journal}{\textit {ibid.}},
  \textbf{\bibinfo{volume}{86}}, \bibinfo{pages}{2854}
  (\bibinfo{year}{2001}{\natexlab{b}});
  \bibinfo{author}{\bibfnamefont{P.~S.} \bibnamefont{Cornaglia}}
  \bibnamefont{and} \bibinfo{author}{\bibfnamefont{C.~A.}
  \bibnamefont{Balseiro}}, \bibinfo{journal}{Phys. Rev. B}
  \textbf{\bibinfo{volume}{66}}, \bibinfo{pages}{115303}
  (\bibinfo{year}{2002}{\natexlab{a}});
  \textbf{\bibinfo{volume}{66}}, \bibinfo{pages}{174404}
  (\bibinfo{year}{2002}{\natexlab{b}}).


\bibitem[{\citenamefont{MacKinnon and Kramer}(1981)}]{Mackinnon1981}
\bibinfo{author}{\bibfnamefont{A.}~\bibnamefont{MacKinnon}} \bibnamefont{and}
  \bibinfo{author}{\bibfnamefont{B.}~\bibnamefont{Kramer}},
  \bibinfo{journal}{Phys. Rev. Lett.} \textbf{\bibinfo{volume}{47}},
  \bibinfo{pages}{1546} (\bibinfo{year}{1981}).

\bibitem[{\citenamefont{Dobrosavljevi{\'c} and
  Kotliar}(1997)}]{Dobrosavljevic1997}
\bibinfo{author}{\bibfnamefont{V.}~\bibnamefont{Dobrosavljevi{\'c}}}
  \bibnamefont{and} \bibinfo{author}{\bibfnamefont{G.}~\bibnamefont{Kotliar}},
  \bibinfo{journal}{Phys. Rev. Lett.} \textbf{\bibinfo{volume}{78}},
  \bibinfo{pages}{3943} (\bibinfo{year}{1997}).

\bibitem[{\citenamefont{Miranda and Dobrosavljevic}(2001)}]{Miranda2001}
\bibinfo{author}{\bibfnamefont{E.}~\bibnamefont{Miranda}} \bibnamefont{and}
  \bibinfo{author}{\bibfnamefont{V.}~\bibnamefont{Dobrosavljevic}},
  \bibinfo{journal}{Phys. Rev. Lett.} \textbf{\bibinfo{volume}{86}},
  \bibinfo{pages}{264} (\bibinfo{year}{2001}).


\bibitem[{\citenamefont{Hirsch}(1992)}]{Hirsch1992}
\bibinfo{author}{\bibfnamefont{M.~J.}~\bibnamefont{Hirsch}},
  \bibinfo{author}{\bibfnamefont{D.~F.}~\bibnamefont{Holcomb}},
  \bibinfo{author}{\bibfnamefont{R.~N.}~\bibnamefont{Bhatt}},  \bibnamefont{and}
  \bibinfo{author}{\bibfnamefont{M.~A.}~\bibnamefont{Paalanen}},
 \bibinfo{journal}{Phys. Rev. Lett.} \textbf{\bibinfo{volume}{68}},
  \bibinfo{pages}{1418} (\bibinfo{year}{1992}).

\bibitem{Ketteman2005}
\bibinfo{author}{\bibfnamefont{S.}~\bibnamefont{Kettemann}} \bibnamefont{and}
\bibinfo{author}{\bibfnamefont{E.~R.}~\bibnamefont{Mucciolo}},
 \bibinfo{journal}{Pis'ma Zh. Eksp. Teor. Fiz.} \textbf{\bibinfo{volume}{83}},
  \bibinfo{pages}{284} (\bibinfo{year}{2006})
 \bibinfo{journal}{[JETP Lett. (to be published)]}.
\end{thebibliography}
\end{document}